 \documentclass[multi]{EngC} 

  \usepackage[rightcaption,raggedright]{sidecap}
  \usepackage{framed}         
  \usepackage{soul}           

  \usepackage[agsm]{harvard}

  \usepackage{rotating}
  \usepackage{floatpag}
  \rotfloatpagestyle{empty}

  \usepackage{amsthm}
  \usepackage{graphicx}

  \usepackage{multind}
  \makeindex{authors}
  \makeindex{subject}

\begin{document}



  \mainmatter



  \alphafootnotes
  \author[Andr\'e E.X. Brown and William R. Schafer]
    {Andr\'e E.X. Brown and William R. Schafer\\
    Medical Research Council Laboratory of Molecular Biology\\
    abrown@mrc-lmb.cam.ac.uk, wschafer@mrc-lmb.cam.ac.uk\\}

  \chapter[Behavioural fingerprinting]{Automated behavioural fingerprinting of \textit{C.~elegans} mutants}

  \arabicfootnotes

  \contributor{Andr\'e E.X. Brown
    \affiliation{Medical Research Council Laboratory of Molecular Biology,
      Hills Road, Cambridge CB2 0QH\\
      e-mail: abrown@mrc-lmb.cam.ac.uk}}

  \contributor{William R. Schafer
    \affiliation{Medical Research Council Laboratory of Molecular Biology,
      Hills Road, Cambridge CB2 0QH\\
      e-mail: wschafer@mrc-lmb.cam.ac.uk}}

\vspace{0.5in}

\section{Introduction}

Rapid advances in genetics, genomics, and imaging have given insight into the molecular and cellular basis of behaviour in a variety of model organisms with unprecedented detail and scope.  It is increasingly routine to isolate behavioural mutants, clone and characterise mutant genes, and discern the molecular and neural basis for a behavioural phenotype.  Conversely, reverse genetic approaches have made it possible to straightforwardly identify genes of interest in whole-genome sequences and generate mutants that can be subjected to phenotypic analysis.  In this latter approach, it is the phenotyping that presents the major bottleneck;  when it comes to connecting phenotype to genotype in freely behaving animals, analysis of behaviour itself remains superficial and time consuming.  However, many proof-of-principle studies of automated behavioural analysis over the last decade have poised the field on the verge of exciting developments that promise to begin closing this gap.

In the broadest sense, our goal in this chapter is to explore what we can learn about the genes involved in neural function by carefully observing behaviour.  This approach is rooted in model organism genetics but shares ideas with ethology and neuroscience, as well as computer vision and bioinformatics. After introducing \textit{C.~elegans} as a model, we will survey the research that has led to the current state of the art in worm behavioural phenotyping and present current research that is transforming our approach to behavioural genetics.

\subsection{The worm as a model organism}

\begin{figure}
    \includegraphics[width=\textwidth]{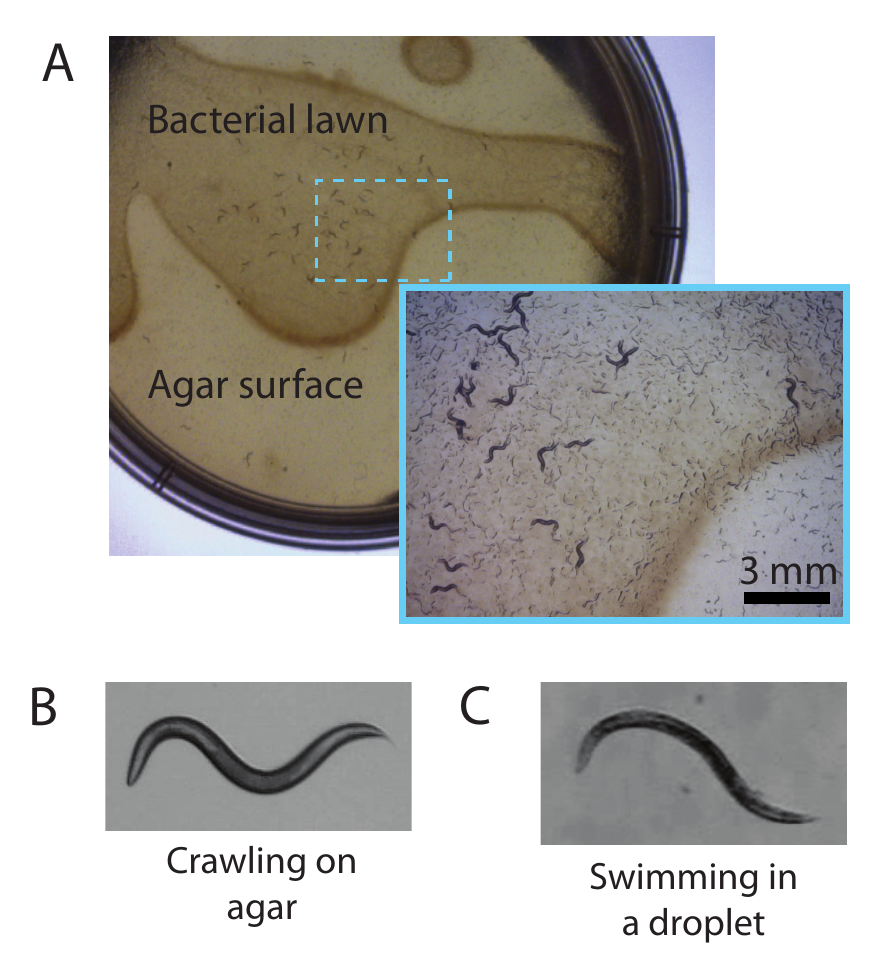}
      \caption{A)  \textit{C.~elegans} is typically cultured on agar plates seeded with bacteria as a food source.  Because of their small size, hundreds of worms can be grown on a standard 5 cm petri dish.  B)  Close-up of a worm crawling on an agar plate.  C)  When placed in a drop of liquid, \textit{C.~elegans} starts to swim.  Swimming is characterised by more rapid body bending with a longer wavelength.}
    \label{crawl-swim}
\end{figure}

\textit{C.~elegans} is a nematode worm that lives in bacteria-rich environments such as rotting fruit and has also been isolated from insects and snails which it is thought to use for longer-range transportation \cite{barriere_high_2005} \cite{lee_nictation_2011}.  In the lab, it is commonly cultured on the surface of agar plates seeded with a lawn of the bacterium \textit{E. coli} as a food source.  On plates, worms lie on either their left or right side and crawl by propagating a sinuous dorso-ventral wave from head to tail.  When immersed in a liquid, they can also swim (Fig.~\ref{crawl-swim}).

As a model organism for the genetics of development, neuroscience, and behaviour, \textit{C.~elegans} offers several advantages.  It progresses through its four larval stages in three and half days with each animal yielding about 300 progeny.  Hermaphrodites can reproduce either by cross- or self-fertilisation; thus, mating behaviour is not essential for the viability of mutant strains.  Indeed, paralyzed animals with almost no nervous system function can be propagated in the laboratory, allowing the analysis of mutants defective in fundamental neuronal molecules such as synaptic vesicle proteins and voltage-gated channels.  Both molecularly and anatomically, \textit{C.~elegans} is exceptionally well-characterised.  It was also the first multicellular organism to have its genome sequenced \cite{sequencing_consortium_genome_1998} and its well-annotated genome as well as information on mutants is available on the community website WormBase.org \cite{yook_wormbase_2012}.  It remains the only organism whose connectome (the wiring diagram of all its 302 neurons) has been completely mapped at electron microscopic resolution \cite{white_structure_1986}.  The connectome and many other details of the worm's anatomy and physiology are available on WormAtlas.org.  Finally, the developmental lineage of each of its 959 adult cells is known and is highly repeatable from one individual to the next \cite{sulston_post-embryonic_1977}.  In other words, given a cell in the adult, one can look up the complete list of divisions that led to that cell from the single-celled embryo.

\subsection{The Genetics of \textit{C.~elegans}}

The genetic study of \textit{C.~elegans} began in earnest with Brenner's 1974 paper in which he screened chemically mutagenised populations of worms for morphological and locomotory defects \cite{brenner_genetics_1974}.  Males occur spontaneously at low rates ($\sim$0.1\%) and this rate can be increased using heat shock.  These males can then be used in genetic crosses and to perform complementation tests and mapping.  Using this approach, Brenner isolated numerous mutants with visible phenotypes, including mutants with movement defects comprising 77 complementation groups.  Since then, many more mutants have been identified, cloned, and often further characterised molecularly using variations on this approach.

Forward genetic approaches continue to play an important role in uncovering the function of nervous system genes, but with the sequencing of the \textit{C.~elegans} genome in 1998 \cite{sequencing_consortium_genome_1998}, reverse genetics is also being extensively used.

Targeted gene deletion has been demonstrated in worms \cite{frokjaer-jensen_targeted_2010}, but has not yet been widely adopted.  In practice, reverse genetics in \textit{C.~elegans} is done principally in two ways: by screening libraries of mutagenised genomes for those containing a deletion of interest and by feeding worms with bacteria that express an appropriate small interfering RNA.

In the first approach, a population of worms is mutagenised using psoralen and UV irradiation which produces more small and medium deletions than the more standard mutagen EMS.  The population is screened using PCR primers designed around the gene to be deleted until a match is found.  Including library construction, a strain carrying a desired deletion allele can be isolated in about two months \cite{ahringer_reverse_2006}.  The procedure has been standardized and is now performed by the \textit{C.~elegans} Gene Knockout Consortium as a service to the worm research community.  Knockouts of new genes, or new alleles of previously knocked out genes, can be ordered online. Knock out and other strains can be ordered for a nominal fee from the \textit{C.~elegans} Genetics Centre at the University of Minnesota.

RNA interference-based gene knockdown can be achieved in \textit{C.~elegans} simply by feeding worms with bacteria that have been modified to express double stranded RNA complementary to the gene to be knocked down \cite{fraser_functional_2000}.  This technique works because most nematode cells express the SID-1 transporter, which enables systemic uptake of double-stranded RNA triggers for RNAi.  Libraries expressing dsRNA fragments of nearly all \textit{C.~elegans} open reading frames have been used to conduct near genome-wide knockdown screens for developmental and morphological phenotypes \cite{kamath_systematic_2003}.  Unfortunately, neurons do not express SID-1, which has limited the applicability of the original method to studying behaviour and nervous system function.  Several approaches have been attempted to increase the efficiency of feeding RNAi with varying degrees of success \cite{simmer_loss_2002} \cite{lehner_loss_2006}  \cite{kennedy_conserved_2004}.  Recently, Calixto \textit{et al}. demonstrated robust feeding RNAi using worms engineered to express SID-1 transgenically in specific groups of neurons \cite{calixto_enhanced_2010}.  This exciting development may prove to be a key enabling technology for a genome-wide investigation of genes affecting behaviour.

With these techniques, as well as the promise of population genomics, the tools for perturbation studies in \textit{C.~elegans} currently significantly outstrip behavioural phenotyping methods in both throughput and sophistication.  However, this is changing, as we will describe after a brief introduction to some of what is already known about \textit{C.~elegans} behaviour.

\subsection{Classic approaches to behaviour}

Nematode behaviour has been studied by researchers for well over 50 years.  Early work examined diverse species and their entire range of behaviour from escape from the egg, through molting, feeding, and mating.  Several modes of locomotion were observed from crawling on solid surfaces to swimming and even jumping.  Croll thoroughly reviewed this early work in \citeasnoun{croll_behavioural_1975}.  For a more current survey of work on nematode behaviour, see \citeasnoun{gaugler_nematode_2004}.

\begin{figure}
    \includegraphics[width=\textwidth]{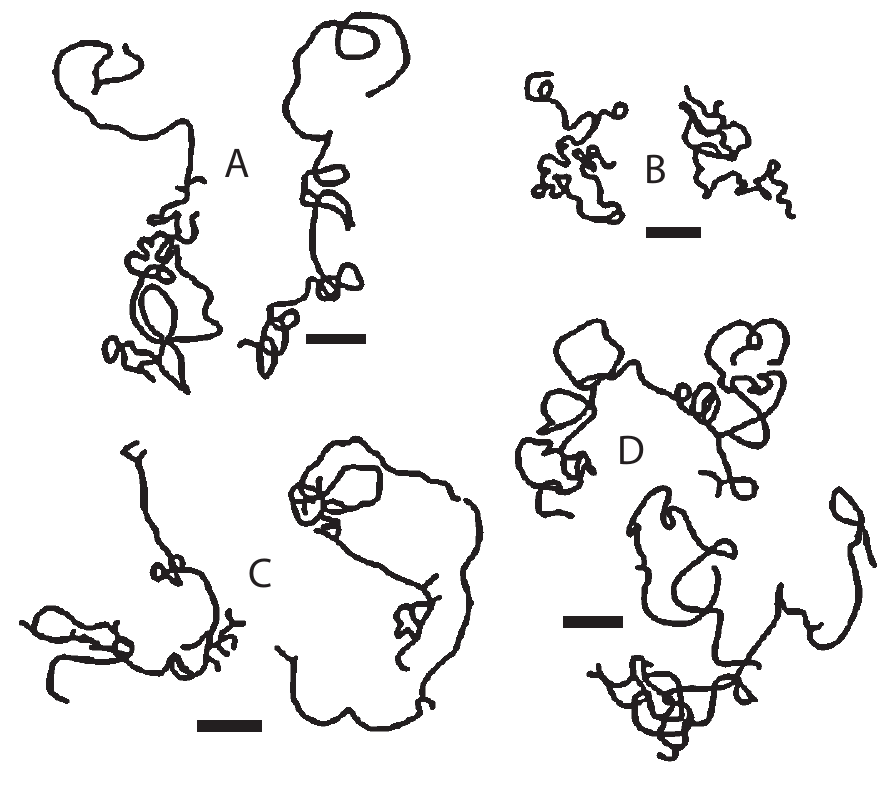}
      \caption{Worms leave tracks in an agar plate and this was used for track analysis.  Adapted from \protect\citeasnoun{croll_components_1975}.  Pairs of tracks labeled A-D are taken from the same individual at different times to illustrate the observed variation.}
    \label{tracks}
\end{figure}

Furthermore, much of this work was not simply descriptive.  Quantitative analysis and modeling was done using computer simulations of aspects of nematode exploration \cite{croll_inherent_1973}, mechanical models of swimming motion \cite{wallace_dynamics_1968}, and quantification of many worm locomotion features including wavelength, frequency and duration of reversals \cite{croll_components_1975}.  These are all still areas of active interest.  Perhaps most relevant to our discussion here is the work on Òtrack analysisÓ in \textit{C.~elegans} in which single or multiple worms are allowed to crawl on a fresh agar surface and inscribe a small crevice.  At the end of the observation time, the agar is transferred to a photographic enlarger and projected onto film.  This provided a permanent record of the data and made subsequent analysis easier (Fig.~\ref{tracks}).

Track analysis gave quantitative insight into possible dispersal strategies and the behavioural mechanisms of chemotaxis.   Interestingly, this method is amenable to high-throughput experiments with very low equipment cost, although as far as we know this has never been attempted.  Many worms could be ÒtrackedÓ simultaneously, limited only by time to transfer single worms to plates and the analysis could be done on a single desktop computer.  Nonetheless, track analysis does have its disadvantages. For example, worms can only be followed until they reach the edge of their arena and there is no information about the dynamics of the motion other than the total time taken for the path.

\section{High Throughput Data Collection and Information Extraction}

There are two classes of problems that must be solved to bring behavioural phenotyping to a level that is truly commensurate with that of current molecular tools.  The first is collecting rich high throughput data and the second is extracting meaning from this wealth of data in an unbiased and quantitative way.  There are several important advantages to both of these aspects of automated behavioural fingerprinting.  The first is simply that it is essential to collect data from a large number of strains to get a sense for the phenotypic landscape and to cover as many interesting genes as possible.  The second is that a carefully engineered data collection pipeline will be naturally standardised and should be more reproducible over time, between operators, and between labs.  Perhaps most importantly, quantitative feature extraction and data analysis lead to abstraction.  Once behavioural data have been abstracted, the precise source of the data matters less than its form, meaning that the extensive tools of statistics and especially bioinformatics can be brought to bear, potentially revealing subtle relationships between phenotypes that may reflect underlying genetic connections.

As we will see in the next two sections many of these problems have been solved at the demonstration level, but the real power of this approach to behavioural genetics has yet to be realised.

\subsection{Worm Tracking: Throughput and Resolution}

\begin{figure}
    \includegraphics[width=\textwidth]{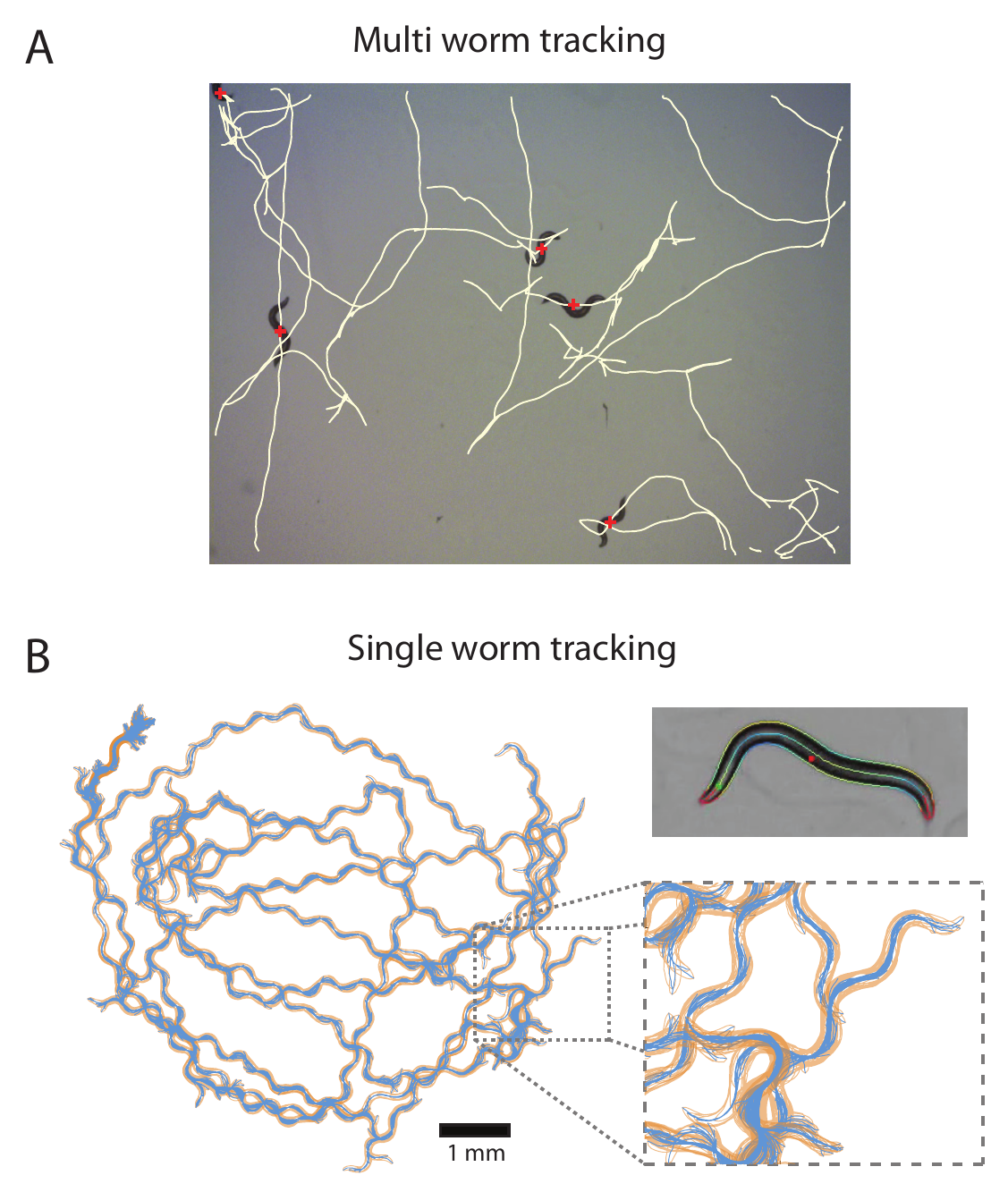}
      \caption{A)  Multiple worms are simultaneously tracked at low resolution using a multi worm tracker.  The results shown here were generated using the Parallel Worm Tracker from the Goodman lab.  B)  A single worm can be tracked using a motorized stage to keep the worm in the camera's field of view.  The worm's outline and skeleton are determined in each frame.  By recording the distance travelled by the stage, a detailed track can be reconstructed including high-resolution posture information (outlines and skeletons from each frame are overlaid here).}
    \label{multiworm-singleworm}
\end{figure}

Worm tracking simply refers to following a moving worm over time, typically recorded in video.  The approaches taken so far in the field can be divided according to resolution: low-resolution tracking allows many worms to be captured in a single field of view (multi-worm tracking) but results in a less detailed picture of each worm's behaviour (Fig.~\ref{multiworm-singleworm}A), while high-resolution single worm tracking provides more detail (Fig.~\ref{multiworm-singleworm}B) but requires the system to follow a single worm in the camera's field of view for long enough to collect a reasonable amount of behaviour.

The first automated worm tracker was described in 1985 in a remarkable paper by Dusenbery \cite{dusenbery_using_1985}.  Using computer and video equipment that are rudimentary by today's standards he was able to record the velocity and reversal rate of 25 worms at 1 Hz, updated in real time.  He and collaborators then used the system to study \textit{C.~elegans}' response to oxygen and carbon dioxide \cite{dunsenbery_video_1985} and to a variety of chemicals \cite{williams_promising_1990}.  Following this pioneering effort, there was a period with relatively little work on automated tracking even though there was rapid progress in behavioural genetics in \textit{C.~elegans} during the same period that could have benefited from the technology.
This eventually changed as more groups discovered the utility of automated behavioural analysis and numerous groups used some kind tracking to investigate worm locomotion, including speed to detect variability in wild isolates from different regions \cite{de_bono_natural_1998}, reversals and turns to understand chemotaxis \cite{pierce-shimomura_fundamental_1999}, or simply using tracking to record long periods of behaviour for subsequent manual annotation \cite{waggoner_control_1998}.  Since then there has been significant interest in developing more user friendly and/or standardized approaches to tracking and analysis both for multi-worm and single worm trackers.

We are aware of three general-purpose multi-worm trackers designed for broad use.  NeMo tracks worm locomotion and includes a graphical user interface that allows users to adjust tracking parameters and correct errors \cite{tsibidis_nemo:_2007}.  It also includes functionality for the subsequent analysis of behavioural data.  The Parallel Worm Tracker uses a simple threshold to distinguish worms from background and works with bright field and dark field videos \cite{ramot_parallel_2008}.  This system also includes a graphical user interface for rejecting or splitting bad tracks and outputs a variety of movement parameters for subsequent analysis. Because both groups make their source code (written in Matlab) freely available, users can add new locomotion metrics relatively straightforwardly.  The most sophisticated multi-worm tracker to date, specifically designed for high throughput, uses Labview to capture data from a 4 megapixel camera as well as to do the first stage of data processing in real time \cite{swierczek_high-throughput_2011}.

The first single worm trackers were introduced to allow higher resolution tracking of an individual for long periods of time; these were used for subsequent manual analysis of egg-laying behaviour  \cite{waggoner_control_1998} or for automated analysis of location data \cite{pierce-shimomura_fundamental_1999} \cite{hardaker_serotonin_2001}.  Because only a single worm is tracked at a time, this approach is more time consuming than multi-worm tracking, but it provides a more nuanced picture of behaviour and makes it possible to extract detailed features of worm shape and locomotion, which we will discuss in the next section.  Single worm trackers typically consist of a motorized stage mounted on a dissecting microscope with a camera.  A computer records video data from the camera that is stored for later analysis but is also used to identify the worm and update the stage position in real time to keep the worm centered in the field of view.  Although there are several labs using single-worm trackers \cite{baek_using_2002}  \cite{stephens_dimensionality_2008} \cite{wang_automated_2009} \cite{feng_imaging_2004}  \cite{cronin_automated_2006}, they have not yet been widely adopted due in part to the expense of the systems and the expertise required to configure the hardware and write the tracking software.

To address this, our group has developed a user-friendly single worm tracker built around a small inexpensive USB microscope.  The entire system can be purchased for under \$5000 USD and will soon be paired with a feature-rich analysis package.  A parts list, instructions for set up, and the tracking software itself can be downloaded from the worm tracker website of the MRC Lab of Molecular Biology (http://www.mrc-lmb.cam.ac.uk/wormtracker/).  Because the system is relatively inexpensive, we have been using it to run eight single worm trackers in parallel and thus increase the throughput of single worm tracking.

\begin{figure}
    \includegraphics[width=\textwidth]{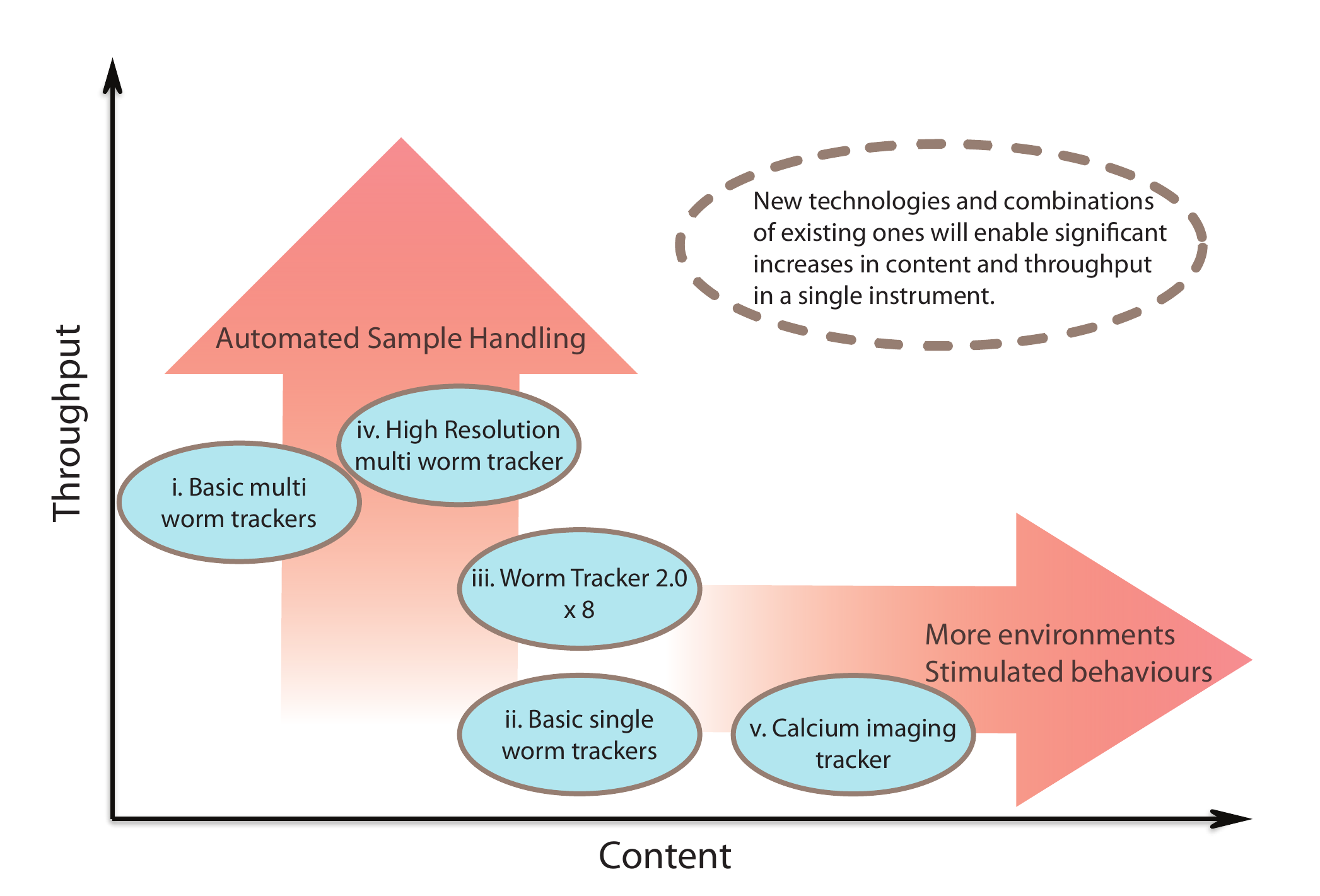}
      \caption{Schematic illustration of the current tradeoff between content and throughput in behavioural phenotyping.  Integrating automated sample handling will significantly improve throughput for imaging systems that can run sufficiently autonomously because picking worms to plates is time consuming and labour intensive.  Content can be significantly increased using better imaging methods, especially functional neural imaging.  In the near future it should be possible to extend both content and throughput by adapting and integrating existing technology.}
    \label{throughput-resolution}
\end{figure}

This highlights the basic trade-off between single and multi-worm tracking---throughput versus resolution---and how these limitations are being addressed from both sides (Fig.~\ref{throughput-resolution}).  In the case of multi-worm tracking, higher resolution cameras are making it possible to extract skeleton-based features of worms even in a field of view with many individuals \cite{swierczek_high-throughput_2011}, while less expensive hardware coupled with free software is making it possible to increase the throughput of the inherently high resolution single worm approach.  Still, with current technology, multi-worm trackers can collect data from a larger number of worms more quickly and single worm trackers still have a resolution advantage so the two methods will likely co-exist for some time.

\subsection{Data Reduction and Abstraction: Segment, Skeletonise, Featurise}

Some early applications of worm tracking were simply for data collection with the analysis still done manually from recorded video.  To really take advantage of the wealth of data provided by trackers, it is necessary to develop algorithms to extract worms from background and to ultimately convert the data into a form that is interpretable by humans to give insight and guide future experiments.

Using appropriate lighting conditions \cite{yemini_illumination_2011} it is usually possible to achieve sufficient contrast, even with worms on a bacterial food lawn, that a simple global threshold does a good job of identifying worms.  For multi-worm trackers, this is sometimes coupled with a size range \cite{ramot_parallel_2008} to eliminate large or small background features that are included by the threshold.  For single worm tracking, taking the largest connected component in the thresholded image is often sufficient.  Still, there are circumstances in which more robust approaches are helpful.  For example, worms in more naturalistic environments, in microfluidic devices, in thick food, or interacting with each other present challenges for the simplest thresholding approaches.

In an effort to make a more universal system for worm identification, Sznitman \textit{et al}. have taken a multi-environment model estimation approach to worm segmentation \cite{sznitman_multi-environment_2010}.  In the first frame of the video only, users must identify the worm to train a Gaussian mixture model used to classify background and foreground (worm) pixels.  This model is then used in subsequent frames to identify the worm which is extracted for further analysis.  The same system was able to extract swimming and crawling worms and worms in microfluidic devices \cite{sznitman_multi-environment_2010}.

In addition to the pixel intensities within a frame, there is also significant information in the correlation between frames.  This has been exploited in worm segmentation using a Kalman filter \cite{fontaine_automated_2006} or recursive Bayesian filter \cite{roussel_computational_2007} coupled with a worm model.  Both approaches allow for the separation of overlapping worms which could improve multi-worm trackers which often simply drop worms that are touching each other from the analysis \cite{ramot_parallel_2008}.  It could also help track worms in, for example, a thick bacterial lawn that can sometimes obscure portions of the worm.  Fontaine \textit{et al}. illustrated their method by studying mating \cite{fontaine_automated_2006}, a behaviour that naturally requires partially overlapping individuals!

A series of outlines over time is still not particularly useful without further processing.  For lower resolution multi-worm trackers, the next step in analysis is typically to quantify aspects of locomotion and posture based on the segmentation.  For example, velocity is simply the change in centroid position of each blob over time.  Reversals and turns can be identified as sharp angle changes in the worm's path.  Worm posture can be roughly approximated by the eccentricity of the worm's equivalent ellipse.  For higher resolution multi-worm and single worm trackers, more features are readily calculated.  In addition to morphology measures based on the outline, skeletonisation leads to a further data reduction without loss of relevant information because worms' width profiles are essentially constant over time.

There are several algorithms that are used for skeletonisation.  The first and still widely used computes the skeleton by thinning the thresholded worm to a single pixel and pruning branches based, for example, on taking the longest path through the structure.  This works reasonably well, but because of the possible ambiguity of choosing the correct branch at the ends, it is not ideal for picking up more subtle head-foraging motions \cite{huang_automated_2008}.  It is also possible to use a chamfer distance transform coupled with curvature to estimate a skeleton \cite{sznitman_multi-environment_2010}.  An attractive approach given its speed and simplicity has recently been reported based on finding the points of highest curvature on the outline and tracing the midline connecting these points \cite{leifer_optogenetic_2011}.  Because the tip of the head is centered on a local curvature maximum, it does a good job of picking up subtle head motions.  At the resolution of a typical single worm tracker, worm skeletons are on the order of 100 pixels long.  This is almost certainly an over sampling given that \textit{C.~elegans} has 22 rows of body wall muscles and therefore significantly fewer degrees of freedom than 100, even including the head which is capable of more complicated motions.

Skeleton curvature can be used to identify even subtle differences in body posture and changes in skeleton curvature over time should in principle be able to distinguish different kinds of uncoordination.  Furthermore, looking at specific sequences of postures has allowed the detection of known behaviours including reversals and reorientations called omega turns \cite{huang_machine_2006}.

A complementary approach to \textit{C.~elegans} behavioural representation has been described by Stephens \textit{et al}. \cite{stephens_dimensionality_2008}. They segment and skeletonise the worm in each video frame, but instead of looking at the skeleton positions they analyse skeleton angles, rotated by the mean angle, yielding a position and orientation independent representation of body postures over time.  They found that the covariance matrix of these angles has a relatively smooth structure and that just four eigenvectors can capture 95\% of the shape variance of worms crawling off food.  They call these four principal components eigenworms, which are essentially four basis shapes that can be added together to reconstruct worm postures.  This compact representation led to interesting results on dynamical models of \textit{C.~elegans} and even to the emergence of stereotyped behaviour without the requirement of a central pattern generator \cite{stephens_emergence_2011}.  More recently, we have shown that the eigenworms derived from wild-type animals can also be used to capture the postures of mutant worms, even those that are highly uncoordinated \cite{Brown24122012}.  This extends the applicability of the eigenworm representation and provides a common and compact basis for capturing worm postures.

\subsection{Towards an OpenWorm Analysis Platform}

Given the variety of systems that have been developed for the automated analysis of \textit{C.~elegans} behaviour from videos, we argue that now is a reasonable time to consider coordinating the efforts of individual groups and creating an open platform for worm behaviour.  In part this could consist simply of shared knowledge of protocols and hardware design.  More importantly though, a set of standard functions available in an easy to use and extendable package would help lower the barrier to entry to new groups and focus researchers with interest and skills on developing useful new analysis tools rather than re-inventing the wheel at the start of each new project.  In the ideal case, the project would look something like ImageJ, the open source image analysis package developed at NIH \cite{abramoff_image_2004}.  It contains many core functions in an easy to use interface but most importantly, it allows the incorporation of plugins and has been adopted by an active group of users and developers.  Of course, the possible user base for such a package will be much less than for ImageJ, but its focus will be correspondingly narrower and hopefully still manageable.

\section{Linking Behaviours and Genes}

\subsection{Insights from quantification}

Sometimes knocking out a gene results in little or no observable behavioural consequence.  In these cases, careful quantification can confirm a phenotype suggested by human observation or even reveal completely new phenotypes.  For example, worms lacking an ion channel called \textit{trpa-1} move well and seem healthy when observed under a dissecting scope \cite{kindt_caenorhabditis_2007}.  However, on closer inspection you might notice an abnormally large head swing (sometimes called foraging motion).  By quantifying the rate of head swings a subtle but reproducible phenotype emerged and this defect in foraging helped direct studies that revealed new aspects of \textit{trpa-1} function \cite{kindt_caenorhabditis_2007}.

A related situation arose in the study of proprioception, or sense of body position, in \textit{C.~elegans}.  Mutant worms lacking an ion channel called \textit{unc-8} were previously reported to have no visible phenotype \cite{park_mutations_1986}, but upon closer inspection were found to have a shallower body bend during locomotion than wild-type worms.  This visual impression was confirmed using track analysis \cite{tavernarakis_unc-8_1997}.  Because of the neural circuits where \textit{unc-8} is expressed and its homology to MEC-4 channels involved in mechanosensation, it was hypothesized to have a role in proprioception and the propagation of the travelling wave worms use for locomotion. Similarly, mutants lacking an ion channel called \textit{trp-4} show a posture defect but still move well.  However, in contrast to \textit{unc-8} animals, \textit{trp-4} worms have deeper body bends than the wild-type \cite{li_c._2006}.  Several experiments now strongly suggest that \textit{trp-4} encodes a mechanosensitive ion channel with a role in proprioception \cite{kang_c._2010}.

\begin{figure}
    \includegraphics[width=\textwidth]{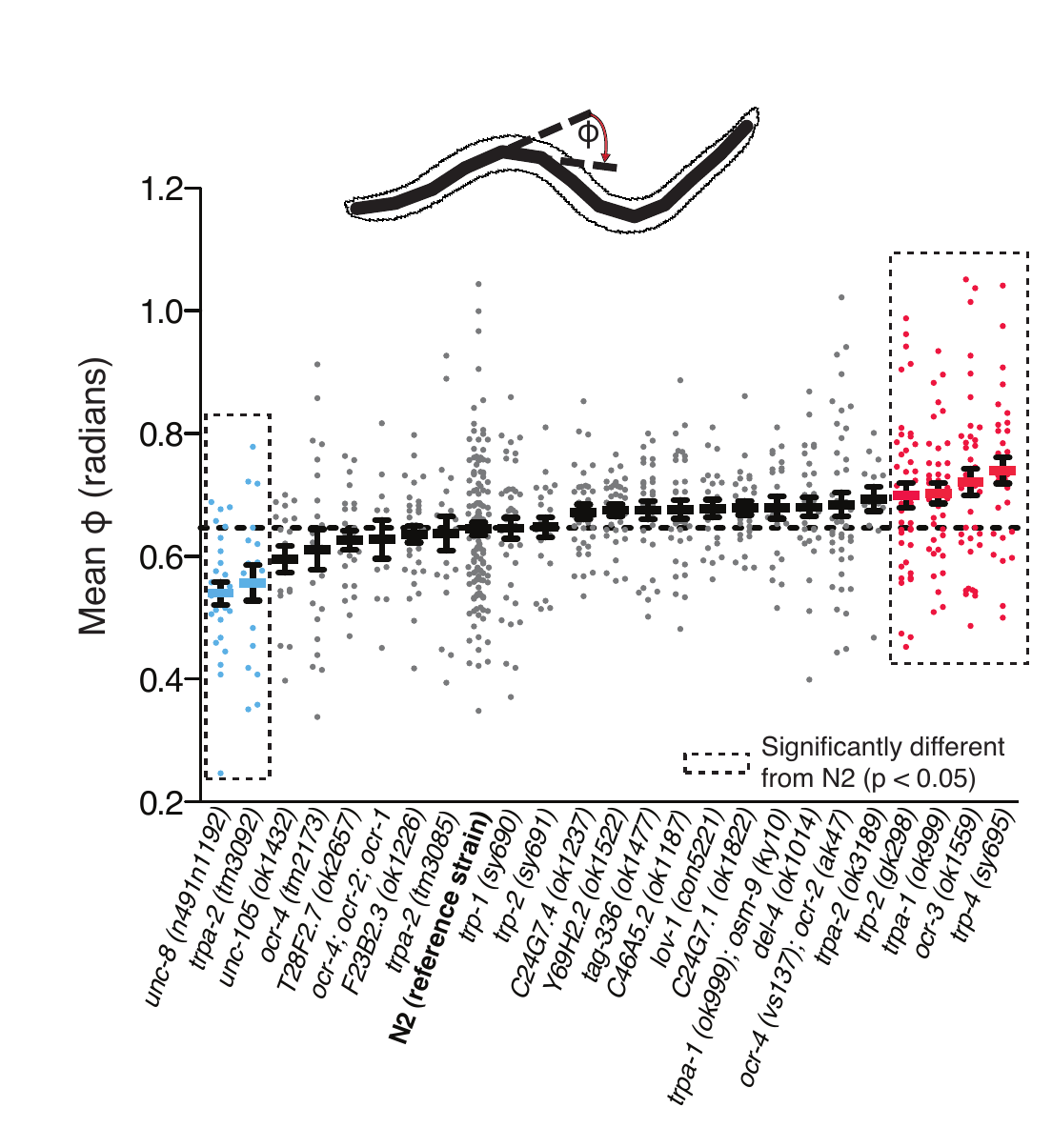}
      \caption{Some mutant strains are more or less curved than the N2 reference strain (wild type).  Each point in the plot represents an individual worm tracked for 15 minutes.  In each frame the change in tangent angle $\phi$ is averaged over the skeleton.  These approximately 25 000 measurements are averaged to give each point shown in the figure.  Black bars are means and error bars show standard errors.  The black dashed line shows the N2 mean.  Strains that are significantly more or less curved than N2 are indicated with the dashed boxes.}
    \label{curvatures}
\end{figure}

There are two aspects of these studies that are particularly interesting.  First, phenotypic quantification was used to confirm subtle phenotypes which then guided functional experiments.  Second, although the phenotypes were subtle, they were picked up by human observers so it was clear what needed to be quantified. To see if there are other channels in the trp and deg channel families (\textit{unc-8} is a DEG/ENaC channel and \textit{trp-4} a TRP channel) that affect curvature, we tracked worms \cite{yemini_preparation_2011} with mutations in one or sometimes several TRP or DEG/ENaC channels.  Several of these mutants were significantly more or less curved than wild-type, possibly suggesting more channels with roles in proprioception (Fig.~\ref{curvatures}).  Note that \textit{unc-8} and \textit{trp-4} are at the extremes of the curvature ranges, perhaps explaining why these channels were the ones initially picked up by human observation. Although the other curvature phenotypes have not been previously reported, quantitative analysis can reliably detect them and they may prove just as useful in guiding functional studies.

\subsection{Unbiased reverse genetic analysis of \textit{C.~elegans} behaviour}

As the curvature example suggests, even looking at a single feature can reveal previously unobserved phenotypes.  What might we learn if we instead took an unbiased look at many mutants and many features?  There have been several papers that demonstrate the potential of this approach.

The first application of automated tracking and machine vision-based analysis to \textit{C.~elegans} was reported by Baek \textit{et al}. in 2002 using single worm tracking and algorithms to extract morphological, locomotion, and posture data \cite{baek_using_2002}.  Using video data recorded from 5 mutant strains, they built a classification tree that could reliably distinguish the different types.  Shortly thereafter, Geng \textit{et al}. used the same system with 8 mutant strains to show that the data could not only be used to classify worms based on their genotypes, but also cluster them based on their phenotypic similarity \cite{geng_automatic_2004}.  Although related, clustering phenotypically similar strains is more relevant than accurately dividing a mixed population into genotype classes because a worm's genotype can be easily and accurately determined directly.  Furthermore, although it may seem reasonable to assume that features with high classification accuracy will also be informative for phenotypic clustering, this is not necessarily the case and even for the small number of strains reported in these papers, somewhat different features were found to be important for the clustering compared to the classification tasks.

Since then, there have been several papers that report automated algorithms to quantify the behaviour of \textit{C.~elegans} mutants.  Without being exhaustive, we will highlight some notable examples.  A similar approach, but with different, complementary features was described by Cronin \textit{et al}. in 2006 \cite{cronin_automated_2006}.  They applied their system to a small number of mutants as a proof-of-concept and showed that they could detect differences in mutant sensitivities to chemicals including arsenite and aldicarb.  More recently, Sznitman \textit{et al}. \cite{sznitman_material_2010} and Krajacic \textit{et al}. \cite{krajacic_biomechanical_2012} have taken a novel approach to feature extraction based on a detailed mechanical model of swimming \textit{C.~elegans} that takes into account both the worm body and its interaction with the surrounding fluid.  What makes their approach unique is that the features they extract---worm stiffness and tissue viscosity---have a direct physical interpretation that cannot be directly seen in the videos.  They show that some mutants with defective muscle can be distinguished from wild-type and from each other based on these estimated parameters.

Stephens \textit{et al}. have used their eigenworm representation combined with stochastic modeling to derive unique metrics that summarise worm behaviour and can distinguish mutant from wild type worms \cite{stephens_emergence_2011}.  As with the approach of Sznitman \textit{\textit{et al}}., these biophysics-based features may serve as useful complements to more empirical measures of behaviour for genetics.  They will be especially useful to the extent that they guide subsequent functional studies by relating a mutant phenotype to specific aspects of a worm's physiology or nervous system.

Multi-worm trackers have also been used to quantify behaviours of mutants and to look at the effects of, for example, drugs \cite{ramot_parallel_2008} and temperature \cite{biron_diacylglycerol_2006}.  As in the case of single worm trackers, these have been used to answer specific phenotypic questions or serve as proof-of-concept experiments on relatively small numbers of mutants. Swierczek \textit{et al}. studied the response of populations of worms to repeated plate tap to find mutants with altered habituation \cite{swierczek_high-throughput_2011}.  They also showed the applicability of their system for studies of chemotaxis. They found some interesting hits in their sample of 33 mutant strains and this will hopefully lead to a larger-scale study.

What all of these studies have in common is their use of human-defined features to represent phenotypes.  However, we do not know if these features are optimal for comparing phenotypes nor whether they are useful for discovering new behaviours.  We have therefore also taken a complementary route, using an unsupervised search for behavioural motifs---short subsequences of repeated behaviour---to define phenotypes \cite{Brown24122012}.  We projected worm skeletons onto wild-type-derived eigenworms and searched for closely repeated subsequences of different lengths and repeated this for many individuals in a large behavioural database.  Some of the motifs represent subtle or irregular behaviours that are nonetheless closely repeated at two different times.  We combined the motifs into a dictionary and each video was then compared to the each of the elements of the dictionary to make a phenotypic fingerprint for that individual.  This is analogous to using features as described above, but now each ``feature'' is a distance from an automatically identified behavioural motif from the dictionary.  Phenotypic comparisons based on the motif dictionary recapitulated some known genetic relationships and were used to hypothesise connections between previously uncharacterised genes \cite{Brown24122012}.

\section{Outlook}

\subsection{Increasing throughput \textit{and} discrimination}

It is clear from the previous sections that the core technologies for high-throughput quantitative phenotyping of \textit{C.~elegans} are within reach, but there remain several challenges that must be overcome.  A framework for understanding the current state of the art and directions for future improvement is summarised in Fig.~\ref{throughput-resolution}.  In essence, it is difficult to achieve both high content and high throughput in a single system.  A multi-worm tracker can collect video data from enough individuals to distinguish at least some phenotypes in about 1 hour.  This number could be decreased with cameras with more pixels and faster computers, bringing, for example, a genome-wide screen within reach.  For groups interested in a particular uncommon phenotype, this could be sufficient.

For example, one of the main phenotypes described by Swierczek \textit{et al}. was tap habituation.  This is the rate at which worms become insensitive to mechanical stimulation provided by repeated taps to their plate.  If there are a relatively small number of genes involved, then a large-scale screen could reveal a tractable number of candidate mutants that can be followed up for further functional analysis.  If, on the other hand, one is interested in proprioception, a genome-wide search for mutants with a curvature phenotype is likely to yield a large number of hits because there are many ways of disrupting neural connections or signaling pathways that lead to an increase or decrease in curvature.  In this case it would be desirable to have a more detailed behavioural fingerprint for clustering worms into related phenotypic classes that may be functionally related as has been demonstrated for a small number of mutants.  Then it would be possible to target follow-up experiments to particular phenotypic sub-classes of particular interest.

This consideration of phenotypic content may be especially critical for future large-scale behavioural screens.  Multi-worm trackers provide greater throughput, but if they lack the sensitivity to meaningfully distinguish or cluster hundreds or thousands of mutant strains this throughput may be difficult to put to good use.  Analysis at this scale has not yet been attempted for worm behaviour so this remains an open question.  The development of high-resolution multi-worm trackers may help \cite{swierczek_high-throughput_2011}.

\subsection{Phenotyping by imaging neuromuscular activity}

Behaviour is the result of a potentially complex feedback between sensation, neuromuscular activity, anatomy, and the physical properties of the environment.  Isolating the contributions from each of these factors is a challenge, but one area where there are likely to be rapid advances is in linking the activity in specific neural circuits and muscles with behaviour and ultimately with genetics.  The main technology driving this advance is the optical recording of neural activity with genetically encoded reporters in freely behaving animals. Fluorescence imaging is relatively non-invasive and is perfectly suited to recording from multiple cells simultaneously. Moreover, nematodes are largely transparent, making their neurons easily accessible to optical recording.  Indeed, the use of genetically-encoded sensors to record neural activity was first demonstrated in \textit{C.~elegans} \cite{kerr_optical_2000}.

Currently the most widely used probes are genetically-encoded calcium indicators (GECIs) that change their emission in response to changes in Ca$^{2+}$ concentration.  This can be either through a change in F\"{o}rster resonance energy transfer (FRET) efficiency in the case of cameleons \cite{miyawaki_fluorescent_1997} or intensity in the case of G-CaMP \cite{nakai_high_2001} and related sensors.  There has also been some recent notable success in genetically encoded sensors designed to directly sense voltage across cell membranes rather than a second messenger like Ca$^{2+}$ \cite{kralj_optical_2011}.  Voltage sensors are not limited by the timescale of cellular Ca$^{2+}$ dynamics and do not rely on the proper functioning of endogenous voltage-gated Ca$^{2+}$ channels. Despite their promise, genetically encoded voltage sensors have not yet been developed to the same extent as Ca$^{2+}$ sensors and are not yet widely used in intact animals.

In worms, GECIs have been used to record the activities of a wide range of neurons.  In immobilised animals, the responses of individual sensory neurons to mechanical \cite{suzuki_vivo_2003}, proprioceptive \cite{li_c._2006}, and chemical \cite{hilliard_vivo_2005}  \cite{chalasani_dissecting_2007} and thermal \cite{biron_diacylglycerol_2006} \cite{kimura_c._2004} stimuli has helped dissect the role of specific neurons in neural circuits \cite{suzuki_functional_2008} \cite{macosko_hub-and-spoke_2009}.  It is this ability to correlate a macroscopic behavioural response with the underlying neural circuits that makes neural imaging such an exciting complement to whole-animal behavioural quantification.  Demonstrations of neural imaging in unconstrained worms are now emerging.  Haspel \textit{et al}. examined the correlation between motor neuron activity and locomotion but were only able to resolve broad activity differences between forward and backward locomotion \cite{haspel_motoneurons_2010}.  This work was extended by Kawano \textit{et al}. who found that an imbalance in motor neuron activity governs the forward or backward state and identified gap junction genes involved in regulating these states \cite{kawano_imbalancing_2011}.  In another study, Faumont \textit{et al}. showed very fast tracking and high-resolution functional imaging using a quadrant photodiode (rather than image) based tracking system \cite{faumont_image-free_2011}.

The combination of neural imaging and worm tracking is in a state analogous to worm tracking itself a decade ago: there have been some interesting demonstrations that have focused on proving feasibility and answering specific questions about locomotion or the action of specific genes.  Genetically encoded Ca$^{2+}$ and voltage indicators have ever-improving sensitivity, response time, and signal to noise ratio \cite{looger_genetically_2012}.  At the same time, sensitive cameras are getting faster and cheaper.  An inexpensive general-purpose system for both tracking and neural imaging would vastly increase the scope of behavioural phenotyping in worms in part because mutations that lead to subtle defects in particular neural circuits may be masked by compensation in another circuit.  This may be a factor that limits the discriminative power in genome-wide screens for behavioural mutants.  Even a coarse-grained view of neural circuit dynamics may offer significant improvements in discriminative power (Fig.~\ref{throughput-resolution}).

Any attempt to integrate neural activity into automated behavioural genetics will face an explosion of data, not just in volume, but also dimension.  Fortunately, with appropriate analysis it should still be possible to devise an abstraction that makes the data amenable to existing (and future) methods for high-dimensional data.  A critical challenge will be to search for a low dimensional representation that still captures important variation as was done for \textit{C.~elegans} locomotion \cite{stephens_dimensionality_2008}.  The generality of this approach \cite{stephens_colloquium_2011} bodes well for applications in neural imaging, especially in a relatively tractable nervous system like \textit{C.~elegans}'.

\subsection{Beyond spontaneous behaviour}

\textit{C.~elegans}' relative simplicity has led to speculation that it may be possible to quantitatively describe its entire behavioural repertoire \cite{stephens_dimensionality_2008}.  This may be possible in the near future for spontaneous locomotion on a surface, but worms are capable of much more.  Exactly how much more is not yet known quantitatively, but finding out will require an extension of behavioural observation beyond spontaneous behaviour on a featureless agar plate.

\textit{C.~elegans} responds to a wide range of stimuli.  For example, when gently touched on their anterior half, worms will reverse.  When touched on their posterior half, they accelerate forward and these responses are controlled by specific neural circuits \cite{chalfie_neural_1985}.  Worms also sense soluble and volatile chemical cues including salt, cAMP, biotin, carbon dioxide, and oxygen \cite{bargmann_chemosensation_2006}.  Worms actively chemotax up attractant gradients and display escape behaviours around noxious chemicals.  Worms also respond to temperature and thermotax toward their most recent cultivation temperature.  They will even accurately track isotherms when placed in a temperature gradient \cite{luo_sensorimotor_2006}.  These behaviours (and the associated genes and neural circuits) have primarily been discovered using manual assays and expert observation, but new technologies are constantly being developed to increase sensitivity and reproducibility and to decrease the subjectivity that can creep into manually scored experiments.  Here we review a selection of promising directions to give a sense for what is currently possible and where it might lead.

If a worm crawling on an agar surface is suddenly placed in a drop of liquid, it will rapidly change to a swimming gait characterised by a higher frequency of body bending and a longer body wavelength (Fig.~\ref{crawl-swim}).  This represents quite an extreme change in the worm's mechanical environment and its response seems to be under genetic control \cite{vidal-gadea_caenorhabditis_2011}.  More subtle transitions are also possible and these have been observed by changing viscosity \cite{korta_mechanosensation_2007} \cite{berri_forward_2009} and more recently with compressive force \cite{lebois_locomotion_2012}.  The advantage of the latter approach is both that the force can easily be changed in real-time and also that there is less chance of inhomogeneity, which can complicate interpretation in studies of swimming in viscous fluids \cite{boyle_gait_2011}.  What makes these studies so interesting is that an accurate response to viscosity will likely require an integration of senses, for example a sense of the applied force of muscles and of touch.  Another possibility would be a sense of applied force combined with sense of posture over time.  The role of touch in gait adaptation has been established \cite{korta_mechanosensation_2007}, but proprioception has not yet been directly implicated and it is not known whether worms have a direct sense of effort. Further tracking experiments combined with genetic and neural ablations as well as careful theoretical work \cite{fang-yen_biomechanical_2010} \cite{sznitman_effects_2010} \cite{ishikawa_integrated_????}  \cite{sauvage_elasto-hydrodynamical_2011} \cite{mailler_biologically_2010-1} will be required to resolve this case of sensory integration and gait computation.

\begin{figure}
    \includegraphics[width=\textwidth]{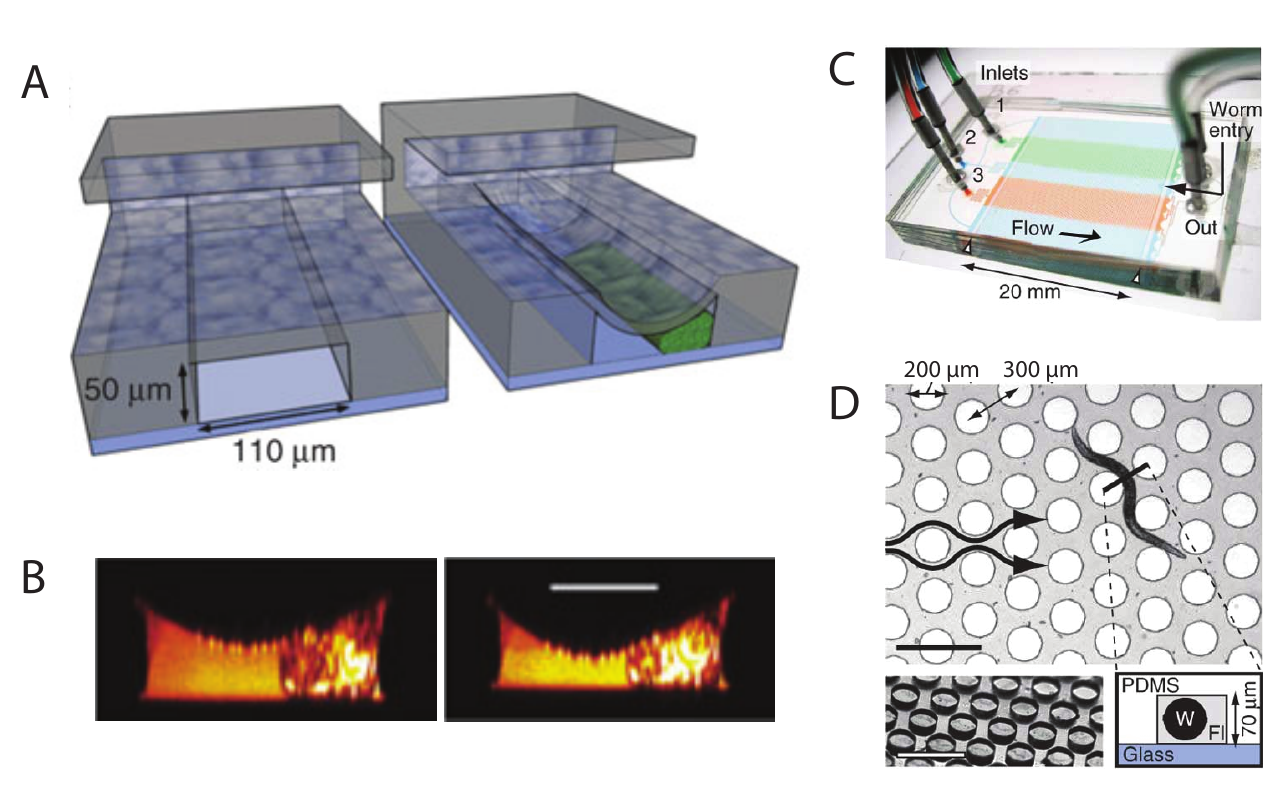}
      \caption{A) Schematic drawing of a pneumatic worm clamping device. B) Fluorescent image of a clamped worm.  A and B are adapted from \protect\citeasnoun{yanik_neurosurgery:_2004}.  C) Microfluidic device for presenting soluble chemicals in sharp steps to worms. D)  Close-up view of a worm in the device shown in C.  C and D are adapted from \protect\citeasnoun{albrecht_high-content_2011}}
    \label{microfluidics}
\end{figure}

Microfluidics has opened many opportunities for precisely handling and immobilizing worms and for providing specific stimulations.  Two-layer devices with pneumatic valves can be used for directing worms to particular chip locations where they can be clamped and assayed (Fig.~\ref{microfluidics}).  This has been used for studies of laser axotomy and axon regeneration \cite{yanik_neurosurgery:_2004}, automated \cite{chung_automated_2009} and semi-automated \cite{hulme_microfabricated_2007} neuron ablation, and for neural imaging of responses to soluble chemicals \cite{chronis_microfluidics_2007} \cite{chokshi_automated_2010}.  See \citeasnoun{chronis_worm_2010} for a review of some of the many related applications.  Of particular interest for extending behavioural phenotyping are applications that control the environment while allowing relatively free locomotion.  Extending the concept of Òarticifcial dirtÓ \cite{lockery_artificial_2008}, Albrecht \textit{et al}. have taken advantage of the laminar flow that exists at low Reynolds number in microfluidic devices to apply precise gradients and sharp boundaries of soluble chemicals to study chemosensation and chemotaxis quantitatively in a potentially high-throughput manner \cite{albrecht_high-content_2011}.  Another particularly simple and provocative experiment suggested that worms could remember the solution to a simple microfluidic T-maze \cite{qin_maze_2007}.

Another method of applying controlled simuli is provided by optogenetics \cite{fenno_development_2011}.  The basic approach is to express a light gated ion channel using promoters specific to a subset of neurons.  Then, when these channels are activated by light they either depolarize and excite or hyperpolarize and inhibit the neurons where they are expressed.  By controlling expression, a simple apparatus can be used to activate or inhibit specific neurons and monitor the effect on behaviour \cite{nagel_light_2005}.  However, because promoters are often not available that are specific to single neurons, it is desirable to be able to target a particular neuron from a population that expresses the light-sensitive channel.  This has the added advantage that different neurons can be targeted in the same animal in a single experiment.  Following from earlier work that combined activation and imaging in immobilized worms \cite{guo_optical_2009}, two groups simultaneously published reports describing the activation of specific neurons in freely behaving animals using a motorized stage and a digital micromirror device \cite{leifer_optogenetic_2011} or a slightly modified projector \cite{stirman_real-time_2011} for local illumination.  The obvious application of these systems is for investigating the role of specific neurons and neural circuits in behaviour, but there will also be applications in genetics.  In particular, even for mutants with no obvious spontaneous locomotion defects it may be possible to detect changes in responses to neuronal stimulation that may reflect more subtle defects in neural circuit function.

\section{Conclusions}

The promise of automated behavioural phenotyping in \textit{C.~elegans} has long been recognized and will soon be realized on a large scale.  Quantitative data enables more meaningful comparisons with models of locomotion that are beginning to emerge and can in some cases lead to new insights into the diversity of---as well as constraints on---behaviour.  The better we understand behaviour the more able we will be to design useful behavioural metrics and to interpret mutant phenotypes.  Already, new algorithms and approaches are showing potential for applications in genetics and with new inexpensive high-throughput methods, open and useable software, and new bioinformatic approaches, these will be more widely adopted and applied to a broad range of topics.

The next decade promises substantial advances in our understand of behaviour and the connection between genotype and this fascinating range of phenotypes.


  \backmatter



  \bibliography{behavioural-fingerprinting}\label{refs}






\end{document}